\def\abs#1{{\left|{#1}\right|}}

\newfont{\elevenmib}{cmmib10 scaled\magstephalf}
\newfont{\ninemib}{cmmib9}
\newcommand{\PRD}[3]{Phys. Rev. {\bf D{#1}} (19{#2}) {#3}}
\newcommand{\PRL}[3]{Phys. Rev. Lett. {\bf {#1}} (19{#2}) {#3}}
\newcommand{\NPB}[3]{Nucl. Phys. {\bf B{#1}} (19{#2}) {#3}}
\newcommand{\PLB}[3]{Phys. Lett. {\bf B{#1}} (19{#2}) {#3}}
\newcommand{\PTP}[3]{Prog. Theor. Phys. {\bf {#1}} (19{#2}) {#3}}

\documentstyle[12pt]{article}
\makeatletter

\@addtoreset{equation}{section}
\makeatother
\begin{document}
\begin{titlepage}
%
%
\vspace{24pt}
\centerline{\Large {\bf{A Cutoff Constrained by the Oblique Parameters}}}
\vspace{10pt}
\centerline{\Large {\bf{ in Electroweak Theory with Two Massless Higgs
Doublets}}}
\vspace{24pt}
\begin{center}
{\bf Kazunori~Takenaga$^{}$\footnote{e-mail:
takenaga@yukawa.kyoto-u.ac.jp,~~JSPS Research Fellow}}
\end{center}
\vskip 0.8 cm
\begin{center}
{\it $^{}$Yukawa Institute for Theoretical Physics,
Kyoto University, Kyoto 606--01 Japan}
\end{center}
\vskip 1.0 cm
\centerline{\bf Abstract}
\vskip 0.2 cm
\baselineskip=15pt
Electroweak theory with two massless Higgs doublets is studied by
solving renormalization group equations for coupling constants
in one-loop approximation.
A cutoff~$\Lambda$, at which one of quartic couplings in the Higgs
potential blows up, is obtained by
imposing constraints from the oblique parameter $T$
on the quartic couplings at low energy.
We find $\Lambda \simeq 0.52\sim 8.4$~TeV at the Higgs mass $M_H=100$ GeV.
The cutoff $\Lambda$ is at most about $60$ TeV
even if we take into account the LEP lower bound
of $M_H\simeq 64$ GeV.
It cannot reach the Planck or GUT scale
due to severe experimental constraints.
It is impossible in the model to realize a large gauge hierarchy
as suggested many years ago by S. Weinberg.
\end{titlepage}
\baselineskip=18pt
\addtolength{\parindent}{2pt}
\setcounter{page}{2}
\setcounter{footnote}{0}
%
\section{\protect{\large\bf Introduction}}
\qquad
It was pointed out by S. Weinberg many years ago\cite{weinberg}
that massless-Higgs-doublet models,
in which radiative corrections induce the spontaneous breakdown of the
$SU(2)\times U(1)$ gauge symmetry ( Coleman-Weinberg mechanism\cite{coleman} ),
may have a possibility to explain the gauge hierarchy.
Even though the quartic couplings of scalar fields defined at
high energy scale, say the Planck ( or GUT ) scale $M_{P(G)}$, are
positive for the stability of theories, they decrease due to
the radiative effects of gauge and Yukawa couplings as energy scale goes down.
New minimum occurs in the potential aside from
the origin by radiative corrections at a
certain low-energy scale, which gives the same order of the VEVs for scalar
fields. An enormous small mass ratio of the weak scale $M_W$
and $M_{P(G)}$ arises as an immediate dynamical
consequence of massless scalar theories coupled to gauge fields.
Yukawa couplings, in general, tend to destabilize the new
minimum. The large Yukawa coupling of $O(1)$ makes the new minimum
unstable in one-massless-Higgs-doublet model, so that
the model is excluded because of the recently announced heavy top
quark mass $M_t\simeq 175$ GeV.
We believe that it is worthwhile to study this attractive possibility
for realizing the gauge hierarchy even in models beyond the minimal
model.
\par
Now the electroweak measurements are so precise that we are at the
stage that we can say something about new physics beyond the minimal
standard model.
The oblique parameters introduced by Peskin and
Takeuchi\cite{takeuchi}
are very useful
and transparent tools for probing or constraining new physics.
In a previous paper\cite{takenaga} we studied the oblique
parameters in electroweak theory with two massless Higgs doublets.
There are four kinds of scalars in the model, charged Higgs
$H^{\pm}$, CP even ( odd ) Higgs $h~( A )$ and scalon $H$.
The scalon is identified with the usual Higgs scalar in the minimal
standard model and its mass is fixed to define the experimental limits
on the oblique parameters.
Hereafter the above Higgs masses are denoted as
$M_{H^{\pm}}, M_h, M_A$ and $M_H$, respectively.
We obtained an allowed region of masses of new particles
by studying the oblique parameter $T$.
A mass relation induced by the Coleman-Weinberg mechanism played an
important and essential role for obtaining it.
The current experiments strictly constrain the mass spectra of new
particles in the model. This means that the quartic couplings in the
Higgs potential, which are written in terms of the masses of new
particles, are also well-constrained.
\par
We think that what Weinberg suggested is very attractive for
realizing the gauge hierarchy
even though we have at present no theoretically reliable
mechanism to guarantee the
masslessness of the Higgs doublets at high energy scale.
It is important to study the possibility of such a large gauge hierarchy in
the two-massless-Higgs-doublets model, taking into account
comprehensively the constraints from the oblique parameters on
coupling constants, in particular, on the quartic couplings in the Higgs
potential and the heavy top quark mass.
It is also interesting to consider what physical pictures
we should make if we cannot get a large gauge hierarchy.
\par
In this paper we solve the RGE's for coupling constants in the
electroweak theory with two massless Higgs doublets in one-loop
approximation\footnote{$\beta$ functions for various couplings in the
model are derived in, for example, \cite{IKN}\cite{FKM}}
to obtain a cutoff $\Lambda$ in the model at which one of the quartic
couplings blows up.
We ignore the effects of the Yukawa couplings except for that of the top
quark in evaluating the RGE's.
Initial conditions for quartic couplings in the equations
are determined from the constraints obtained by the oblique parameter $T$
at reference points $(M_t, M_H)=(175, 100), ~(150, 100), ~(125, 100)$
GeV, where $M_t$ is the top quark mass and $M_H$ is the mass of the
scalon. At each reference point, we obtain allowed mass regions for
new particles.
Picking up some points in the allowed region and
fixing $\tan\beta$ for each point, which is a ratio of the vacuum expectation
values of two Higgs fields, one can completely
fix the quartic couplings at low energy. In our numerical analyses we choose
$\tan\beta=0.6, 1.0, 3.0$.
Using the quartic couplings obtained in this way as the initial
conditions at low energy, we obtain a cutoff $\Lambda$ in the model.
We find that~(i)~the cutoff $\Lambda$ is suppressed if
there are hierarchies among the quartic couplings, while
degeneracies between all the quartic couplings give larger $\Lambda$
{}~(ii)~$\Lambda$'s do not take quite different values among these reference
points, and it is about $\Lambda \simeq 0.52\sim 8.4$ TeV.
{}~(iii)~If $M_H$ becomes smaller it is possible to make $\Lambda$ larger
than these values. It is, however, at most about $60$ TeV if we take into
account
the current experimental limits on the top quark and Higgs masses.
\par
In the next section, we briefly
review the Higgs sector of the electroweak theory with two massless
Higgs doublets. We clarify the relations between masses of new particles
and the quartic couplings in the Higgs potential.
In section $3$ we show the allowed regions of masses of new particles
constrained from the experimental limit on the oblique parameter $T$
at three different reference points.
In the section $4$ we present results of our numerical analyses.
Concluding remarks are given the final section.
\section{\protect{\large\bf The Higgs sector of the model}}
\qquad
The Higgs potential of our model is given by
\begin{eqnarray}
V_H &=&{\lambda_1\over 2}(\Phi_1^{\dagger}\Phi_1)^2+
{\lambda_2\over 2}(\Phi_2^{\dagger}\Phi_2)^2+
\lambda_3(\Phi_1^{\dagger}\Phi_1)(\Phi_2^{\dagger}\Phi_2)+
\lambda_4(\Phi_1^{\dagger}\Phi_2)(\Phi_2^{\dagger}\Phi_1)\nonumber\\
&+&{\lambda_5 \over 2}\Bigl[(\Phi_1^{\dagger}\Phi_2)^2+{\rm
h.c.}\Bigr].
\label{eq:Higgs}
\end{eqnarray}
For the tree-level potential to be stable,
the parameters $\lambda_{1\sim 5}$ must satisfy
\begin{eqnarray}
\lambda_{1, 2}~>~0,
{}~~\lambda_5~&<&~0,~~\lambda_4+\lambda_5~<~0,\label{eq:conda} \\
\sqrt{\lambda_1\lambda_2}&+&\lambda_3+\lambda_4+\lambda_5~\geq~0.\label{eq:condb}
\end{eqnarray}
Here we choose the sign of $\lambda_5$ to be negative.
The third condition in (\ref{eq:conda}) must be satisfied to keep the
$U(1)_{\rm em}$ invariance.
The Higgs potential is a homogeneous polynomial of the Higgs fields so
that the $SU(2)\times U(1)$ gauge symmetry is not broken at the tree level.
One must take into account at least one-loop corrections to the
potential in order to break the symmetry. Detailed discussions on the
one-loop effective potential in the model are
given in \cite{IKN}\cite{FKT}\cite{gildner}.
It is important to note that
the gauge symmetry is broken desirably by radiative
corrections ( Coleman-Weinberg mechanism ) if and only if the
tree-level potential possesses a flat direction realized by the
coupling relation
\begin{equation}
\lambda(M_R)\equiv\sqrt{\lambda_1\lambda_2}+\lambda_3+\lambda_4+\lambda_5=0
\label{eq:flat}
\end{equation}
at some renormalization scale $M_R$.
The flat direction in the VEV space of the Higgs potential
is given by
\begin{eqnarray}
\langle\Phi_1\rangle&=&{1\over \sqrt 2}
\left(\begin{array}{c}0\\v_1\end{array}\right)=
{\rho\over \sqrt 2}
\left(\begin{array}{c}0\\n_{01}\end{array}\right),
\langle\Phi_2\rangle=
{1\over \sqrt 2}
\left(\begin{array}{c}0\\v_2\end{array}\right)=
{\rho\over \sqrt 2}
\left(\begin{array}{c}0\\n_{02}\end{array}\right),\nonumber\\
(\rho &\geq & 0),\nonumber
\end{eqnarray}
where
$$
n_{01}^2={\sqrt\lambda_2 \over {\sqrt\lambda_1 +\sqrt\lambda_2}},~~
n_{02}^2={\sqrt\lambda_1 \over {\sqrt\lambda_1 +\sqrt\lambda_2}}.
$$
The order parameter $\rho^2$ is fixed by
$$
\rho^2=v^2\equiv {1\over {{\sqrt 2}G_F}},
$$
where $G_F$ is the Fermi coupling constant.
The renormalization scale $M_R$ at which (\ref{eq:flat}) is
realized is related to $v$ by
$$
M_R^2={\rm e}^{-11/3}~v^2
$$
through the Coleman-Weinberg mechanism.  \par
The physical Higgs masses are obtained by the quadratic
terms of the potential with respect to fields.
We have
\begin{eqnarray}
M_{H^{\pm}}^2&=&{1\over 2}\bigl(\sqrt{\lambda_1\lambda_2}+\lambda_3
\bigr)v^2,~~
M_A^2=-\lambda_5 v^2,~~M_h^2=\sqrt{\lambda_1\lambda_2}v^2,\label{eq:newmass}\\
M_H^2&=&{G_F\over {4{\sqrt 2}\pi^2}}
\Bigl[6M_W^4+3M_Z^4-12M_t^4+2M_{H^{\pm}}^4+M_h^4+M_A^4\Bigr]\label{eq:scalon}.
\end{eqnarray}
We identify the scalon denoted by $H$, which is the pseudo-Goldstone boson
associated
with the scale invariance of the Higgs potential, as the usual Higgs boson in
the
minimal standard model. Its mass is defined
by $M_H^2\equiv d^2 V_1/d\rho^2|_{\rho=v}$, where $V_1$ is the
one-loop effective potential in our model.
By defining $\tan\beta$ as
\begin{equation}
\tan^2\beta\equiv {v_2^2 \over v_1^2}
={{\sqrt \lambda_1}\over {\sqrt\lambda_2}},
\label{eq:tan}
\end{equation}
the five quartic couplings are written in terms of the masses
of new particles;
\begin{eqnarray}
\lambda_1&=&{M_h^2 \over v^2}\tan^2\beta,
{}~\lambda_2={M_h^2 \over v^2}{1\over \tan^2\beta},\label{eq:lama}\\
\lambda_3&=&{1\over v^2}(2M_{H^{\pm}}^2 - M_h^2),
{}~\lambda_4={1\over v^2}(M_A^2 - 2M_{H^{\pm}}^2),
{}~\lambda_5=-{M_A^2\over v^2}\label{eq:lamb}.
\end{eqnarray}
{}From (\ref{eq:lama})~and (\ref{eq:lamb}) we see that
$\lambda_{3\sim 5}$ are fixed once we fix the masses $M_{i}$
( $i=H^{\pm}, h, A$ ).
The determination of $\tan\beta$ fix $\lambda_1$ and $\lambda_2$.
Note that all the quartic couplings are defined at the energy
scale $M_R$. The masses of new particles are constrained by the
oblique parameter $T$ as we will see in the next section.
\section{\protect{\large\bf Constraints from the oblique parameter $T$}}
\qquad
Let us obtain the allowed mass region for new particles,
$M_{H^{\pm}}, M_A$ and $M_h$ by studying the oblique parameters.
The contributions to the parameters from new particles are studied in
\cite{takenaga} at a reference point ${\rm I}(M_t, M_H)=(175, 100)$ GeV
in which we found that the constraint on the masses
mainly comes from the parameter $T$.
We had almost no constraints on them from the parameter $S$ at the
reference point.
This situation does not changed in
the case for reference points II and III defined below.\par
In order to see the behavior of the cutoff in our model with respect to
the heaviness of the top quark, we choose following reference points;
$$
(M_t, M_H)={\rm I}(175, 100), ~{\rm II}(150, 100), ~
{\rm III}(125, 100)~{\rm GeV}
$$
At each reference point, the experimental limits on $T$ are
calculated to be\cite{barbieri}\cite{ABC}
\begin{eqnarray}
-0.62&\leq& T~ \leq ~-0.14~~{\rm for~~I}   \nonumber \\
-0.32&\leq& T~ \leq ~+0.16~~{\rm for~~II}  \nonumber \\
-0.06&\leq& T~ \leq ~+0.42~~{\rm for~~III}.\label{eq:tvalues}
\end{eqnarray}
In Figs. $1\sim 3$ we display the allowed mass region for $M_{H^{\pm}}$
and $M_h$ by using (\ref{eq:tvalues}). The areas inside the
solid curves are the allowed mass regions
from the $1\sigma$ errors in $T$.
In the previous paper we assumed $M_h \leq M_A$ because the parameter
$T$ is symmetric under $M_h\leftrightarrow M_A$. In this paper,
however, we do not assume any possible mass hierarchies among new
particles to study allowed quartic couplings comprehensively.
At the reference point I, $T$ takes negative values at the $1 \sigma$
errors. A relation $M_h < M_{H^{\pm}}< M_A$ or
$M_A < M_{H^{\pm}}< M_h$ must be satisfied to obtain the negative $T$.
A left ( right ) domain of the allowed regions in Fig. $1$ corresponds to the
case $M_{h(A)} < M_{H^{\pm}}< M_{A(h)}$.
In order for the parameter $T$ to make sense, the masses of new
particles must be larger than the $Z$-boson mass $M_Z$.
We set $M_i~(i=H^{\pm}, h, A)\geq 140$ GeV for illustration.
This limit does not alter our numerical results for the cutoff
significantly.
Note that the values of $M_{H^{\pm}}$ can be determined once we fix
$M_A$ and $M_h$. This is because, from the expression
(\ref{eq:scalon}), one obtains the mass relation
\begin{equation}
M_{H^{\pm}}^2
=\Bigl[{{2{\sqrt 2}\pi^2} \over G_F}M_H^2-{1\over 2}(6M_W^4+3M_Z^4-12M_t^4)
-{1\over 2}(M_A^4+M_h^4)\Bigr]^{1/2}.\label{eq:relation}
\end{equation}
The oblique parameter $T$ depends on three
parameters, $M_{H^{\pm}},~M_h$ and $M_A$, but the
mass relation (\ref{eq:relation}) reduces the number of free
parameters to two, for example, $M_A$ and $M_h$.
\par
For the later analyses we note the custodial symmetry of the Higgs
potential in our model.
The experimental limits on
$T$ include the value of zero in cases of the reference points II and III.
In order to explain this value the Higgs potential must have the
custodial symmetry.
The custodial symmetry in our model can be realized if
$M_{H^{\pm}}=M_A$ or $M_{H^{\pm}}=M_h$\cite{pomarol}.
The former case implies $\lambda_4=\lambda_5$ with $\tan\beta$ being
free, and the later case does
$\lambda_1=\lambda_2=\lambda_3$ with $\tan\beta=1.0$.
The custodial symmetry is not necessary in the case of the reference
point I because $T$ is negative at the $1\sigma$ errors.
We also note that if all the masses of new particles are degenerate we
have $\lambda_1=\lambda_2=\lambda_3=-\lambda_4=-\lambda_5$ for
$\tan\beta=1.0$.
\section{\protect{\large\bf Numerical results}}
\qquad
Let us pick up some points which cover the almost all allowed
region of masses of new particles obtained in the section $3$.
The points we pick up are labeled by A, B, {\em etc} in Figs. $1\sim 3$.
For the right domain of the allowed regions in Fig. $1$, we
take those points which are obtained from the points in the left
domain by an exchange of $M_h$ and $M_A$. This is possible because
both domains are related each other
by this exchange due to the symmetric property of $T$ and $M_{H^{\pm}}$
under $M_h \leftrightarrow M_A$.
At each point we calculate the quartic
couplings $\lambda_{1\sim 5}$ by (\ref{eq:lama})
and (\ref{eq:lamb}) for each $\tan\beta=0.6,~1.0,~3.0$.
We use these couplings as initial conditions of
the RGE's at low energy in our model.
One should be careful that the
conditions (\ref{eq:conda}) and (\ref{eq:condb}) must be satisfied
in evaluating the RGE's, otherwise the
system we concern becomes unstable.
The absolute values of the initial conditions we shall use are within ranges
{}~$0.05 ~<~ |\lambda_i|~<~4.0$ ( $i=1\sim 5$ ).
The quartic couplings constrained by $T$ are large enough
, so that the heavy top quark such as $M_t\sim 175$ GeV does not
destabilize the true minimum of the potential in our model.
This point is quite different from the model with one massless Higgs
doublet.
\par
In Tables $1\sim 3$ we present numerical results of the values of a
cutoff $\Lambda$ for the reference points I $\sim$ III, respectively
\footnote{The points listed in Table $2$ and $3$ are the case with
no custodial symmetry in the Higgs potential. We shall discuss the
case with the symmetry separately.}.
There are the cases with $\Lambda~<~M_{i}$, where $i$ stands for
$H^{\pm}, h$ or $A$. These cases denoted by values in parentheses
in the tables are not acceptable.
We observe that $\Lambda$'s do not take quite different values
among these tables.
This means that the effect of the heavy top quark on $\Lambda$
does not appear seriously at these reference points.
This can be traced back to magnitudes of
the mass of the scalon we choose in the reference points.
It is seen from (\ref{eq:scalon}) that the quartic couplings must
be appropriately large enough in order for the mass of the
scalon to be $100$ GeV for the heavy top quark masses.
The initial values of the quartic couplings are sufficiently
large, so that the running of the couplings is not affected strongly
by the large Yukawa coupling, that is, the heavy top quark.
Therefore, the sensitivity
of the cutoff to the heavy top quark is small and the values of $\Lambda$
is also small at the reference points.
We will discuss the possibilities for obtaining larger cutoff in the
last paragraph of this section.
\par
At each point in the tables the largest cutoff is obtained for
$\tan\beta=1.0$, which means $\lambda_1=\lambda_2$.
A large difference $\lambda_1 \ll(\gg)\lambda_2$ are produced when
$\tan\beta = 0.6$ ( $3.0$ ) ( see (\ref{eq:tan})).
Then, one of the couplings blows up faster than the other coupling.
This is why there are many $\Lambda < M_i$ cases in these two
values of $\tan\beta$.
When $\lambda_1$ or $\lambda_2$ exceeds about $5$, the cutoff
$\Lambda$ lies below one of
mass spectra of new particles as far as our numerical analyses are
concerned.
For example, at the point E ( M ) with $\tan\beta=3.0~( 0.6 )$ in Table
$2$, we have
$(\lambda_1, \lambda_2)\simeq (5.9~( 0.74 ), 0.07~( 5.7 ))$.
\par
We observe that $\Lambda$ is larger when the hierarchies among the
quartic couplings are smaller.
Let us explain this feature by comparing
the points C and L in the Fig. $2$ as an example.
The quartic couplings at the points for $\tan\beta=1.0$
are calculated as
\begin{eqnarray*}
{\rm C}(\lambda_1,\lambda_2,\lambda_3,\lambda_4,\lambda_5)&\simeq&
(0.37,0.37,3.67,-1.54,-2.50),\quad \Lambda\simeq 3.95\quad{\rm TeV},\\
{\rm L}(\lambda_1,\lambda_2,\lambda_3,\lambda_4,\lambda_5)&\simeq&
(1.48,1.48,2.56,1.99,-2.04),\quad \Lambda\simeq 6.41\quad{\rm TeV}.\\
\end{eqnarray*}
{}From these values we obtain
\begin{eqnarray*}
1.2\le \delta\lambda_{ij}&\equiv&
\abs{\abs{\lambda_i}-\abs{\lambda_j}}\le 3.3\quad
{\rm for \quad C},\\
0.52\le \delta\lambda_{ij}&\equiv&
\abs{\abs{\lambda_i}-\abs{\lambda_j}}\le 1.08\quad
{\rm for\quad L},\\
\end{eqnarray*}
where $i~(>~j)$ runs from $1$ to $5$.
The hierarchies among the quartic couplings become smaller and smaller
to give larger cutoff as shown Table $2$ when we move the points along
C$\rightarrow$G$\rightarrow$J$\rightarrow$L$\rightarrow$N.
$\delta\lambda_{ij}$ at the point N takes almost the same values with the
point L. But the absolute values of each quartic coupling at N are
somewhat larger than those of L, so that $\Lambda$ at L becomes slightly
larger than $\Lambda$ at N.
The same tendency for degenerate quartic couplings are also seen in Fig. $2$
by D$\rightarrow$E$\rightarrow$F$\rightarrow$G and
A$\rightarrow$D$\rightarrow$H$\rightarrow$K$\rightarrow$N.
These behaviors for smaller hierarchies among the couplings
are common in three reference points we choose.
\par
The maximum value of the cutoff in Table $1, 2$ and $3$ is obtained
at the point G, L and K, respectively.
The measure of the hierarchies among the quartic couplings defined by
$\delta\bar\lambda \equiv \delta\lambda_{ij}^{\rm max}-\delta\lambda_{ij}^{\rm
min}$ takes the smallest values at these points
compared with other points in each reference point.
In particular the masses of new particles are almost degenerate at
the point L and K.
$\delta\bar\lambda$ for L and K are about $0.56$ and $0.74$,
respectively, but that for G is more than $2$.
This is because
at the reference point I the parameter $T$ is negative at
the $1\sigma$ errors, and it is possible when a relation
$M_{h(A)}~<~M_{H^{\pm}}~<~M_{A(h)}$ is satisfied, which means
that there are still hierarchies among the quartic couplings.
So $\Lambda$ at G is not so large
compared with $\Lambda$'s at each point of L and K.
\par
Let us study the behaviors of the cutoff when the Higgs potential
has the custodial symmetry.
At the points
A, D, H, K, N ( A, D, G, J, N ) in the Fig. $2~(3)$
the custodial symmetry, implying
$M_{H^{\pm}}=M_h$ with $\tan\beta=1$, exists in the Higgs potential.
The other custodial symmetry realized by
$M_{H^{\pm}}=M_A$ with $\tan\beta$ being free
are obtained
by exchanging the values between $M_h$ and $M_A$ at each point.
The results for both cases
are summarized in Tables $4$ and $5$ for the reference points II and
III, respectively.
We find, again, that $\Lambda$ takes larger value as
the hierarchies among the quartic couplings are smaller;~
A$\rightarrow$D$\rightarrow$H$\rightarrow$K$\rightarrow$N~
(~A$\rightarrow$D$\rightarrow$G$\rightarrow$J$\rightarrow$N~).
At the last points N's in the sequences $\delta\bar\lambda$ takes the
smallest values compared with other points.
\par
Now we shall discuss briefly the value of
the cutoff when the mass of $M_H$ becomes small.
We vary the Higgs mass $M_H$ to see how the
values of $\Lambda$ change and its sensitivity to the top quark.
We analyze them at $M_H=80, 60, 40, 20$ GeV for each value of
$M_t=175$ GeV and $125$ GeV.
{}From the previous analyses for the cutoff we found that
$\Lambda$ takes larger value when the masses of new particles are
almost degenerate.
So we assume
$M_{H^{\pm}}=M_h=M_A~(~\equiv M_{\rm new}~)$
with $\tan\beta=1.0$, which means
$\lambda_1=\lambda_2=\lambda_3=-\lambda_4=-\lambda_5$ as mentioned
in the section $3$.
We present results of $\Lambda$ for these cases in Table $6$.
When $M_H$ becomes light, the quartic couplings become smaller
for both $M_t=175$ GeV and $M_t=125$ GeV cases to yield larger cutoff.
In particular, for $M_t=125$ GeV the degrees of decreasing the values of
the quartic couplings are large compared with $M_t=175$ GeV case, so that
the difference of the
cutoff between $M_t=175$ GeV and
$M_t=125$ GeV for each value of $M_H$ is enhanced as $M_H$ becomes
lighter. Apparently, we see the effect of the top quark on $\Lambda$.
If we take $M_H=10$ GeV for $M_t=125$ GeV, which corresponds
to
$$
\lambda_1=\lambda_2=\lambda_3=-\lambda_4=-\lambda_5\simeq
0.45,~M_{\rm new}\simeq 165~{\rm GeV},
$$
$\Lambda$ is about $1.06\times 10^{14}$ GeV.
A large gauge hierarchy can be realized, but the constraints from
the LEP bound of $M_H\simeq 64$ GeV and the announced heavy top quark
mass are not satisfied. The top and Higgs masses are so
heavy that the cutoff cannot reach the Planck or GUT scale.\par
\section{\protect{\large\bf Concluding Remarks}}
\qquad
We have studied the one-loop RGE's in the
electroweak theory with two massless Higgs doublets. Initial
conditions of the equations for the quartic couplings are fixed by the
constraints obtained from the oblique parameter $T$ at three reference
points $(M_t, M_H)=(175, 100), ~(150, 100), ~(125, 100)$ GeV.
The constrained quartic couplings are large enough that the true
minimum in the potential of the model is not destabilized by
the heavy top quark. We have found the cutoff
$\Lambda$ in the model for the reference points.
The result is summarized as $\Lambda\simeq 0.52\sim 8.4$ TeV.
The values of the cutoff among these reference points are not so different.
The cutoff becomes larger when the quartic couplings are
almost degenerate, which also includes the case with the custodial
symmetry in the Higgs potential.
The value of the cutoff is at most about $60$ TeV, and it is impossible
to obtain larger values of $\Lambda$ than this value from the current
experimental limits on the Higgs and the top quark masses.
The heavy top quark and LEP bound on
the Higgs mass prevent $\Lambda$ from reaching the high energy scale.
It is impossible to realize the large
gauge hierarchy as suggested many years ago by S. Weinberg
in the electroweak theory with two massless Higgs
doublets because of the constraints on $\Lambda$
from the oblique parameter $T$
and the heavy top and Higgs masses. \par
The obtained cutoff $\Lambda$ indicates that the physics we have
considered must be changed around $\Lambda$.
It is natural to take a point of view that our model is an effective
theory at low energy of some fundamental dynamics.
When the energy scale approaches
to $\Lambda$, the quartic couplings become large and are outside the
validity of the perturbation theory. This implies that
nonperturbative dynamics governs the physics near the scale $\Lambda$.
It may be possible to consider that the massless Higgs fields are
Nambu-Goldstone-boson fields associated with the breakdown of some symmetries.
This possibility is natural in the usual sense that effective theories
forbid scalar mass terms for naturalness.
One may also consider that the Higgs potential
(\ref{eq:Higgs}) is regarded as a part of the low-energy effective
( Ginsburg-Landau ) lagrangian of
composite Higgs fields, which are bound states of fermion and anti-fermion.
In this case the mass
terms for the composite Higgs fields at the tree level are renormalized to be
zero, and one should take into account the effects of higher
dimensional operators in discussing the low-energy physics because
of a small hierarchy between $v\sim 246$ GeV and $\Lambda$ we have
studied.
\vspace{15pt}
\begin{center}
{\bf Acknowledgment}
\end{center}
The author would like to thank Prof. K.~Inoue and A.~Kakuto for
careful reading of this
manuscript and various comments. This work was
supported by Grant-in-Aid for Scientific Research Fellowship, No.5106.
\newpage

%
%
\vspace{40pt}
{\large\bf Figure Captions}\par\noindent
\begin{itemize}
\item[Fig.1] The allowed mass regions for $M_{H^{\pm}}$ and $M_h$ from the
oblique parameter $T$ at the reference point I. Points denoted by
A$\sim$ H in the region give initial conditions of the
quartic couplings for solving the RGE's. The quartic couplings in the
right domain side of the allowed regions are calculable as explained in
the text.
\item[Fig.2] The allowed mass region for $M_{H^{\pm}}$ and $M_h$ from the
oblique parameter $T$ at the reference point II. Points denoted by
A$\sim$ Q in the region give initial
conditions of the quartic couplings for solving the RGE's.
At the points A, D, H, K and N the Higgs potential has the custodial
symmetry which implies $M_{H^{\pm}}=M_h$ with $\tan\beta=1.0$.
\item[Fig.3] The allowed mass region for $M_{H^{\pm}}$ and $M_h$ from the
oblique parameter $T$ at the reference point III. Points denoted by
A$\sim$ Q in the region give initial
conditions of the quartic couplings for solving the RGE's.
At the points A, D, G, J and N the Higgs potential has the custodial
symmetry which implies $M_{H^{\pm}}=M_h$ with $\tan\beta=1.0$.
\end{itemize}
%
{\large\bf Table Captions}
\par\noindent
\begin{itemize}
\item[Table 1] The cutoff $\Lambda$ obtained at the reference point
$( M_t, M_H )=( 175, 100 )$ GeV for $\tan\beta=0.6,~1.0,~3.0$.
The value of $M_A$ is determined once we fix the point on
$M_{H^{\pm}} - M_h$ plane by the relation (\ref{eq:relation}).
The values of left ( right ) side in the row of $\Lambda$
correspond to results obtained by
quartic couplings in the left ( right ) domain of the allowed regions in
Fig. $1$.
\item[Table 2] The cutoff $\Lambda$ obtained at the reference point II
for $\tan\beta=0.6,~1.0,~3.0$.
The value of $M_A$ is determined once
we fix the point on $M_{H^{\pm}} - M_h$ plane by the relation
(\ref{eq:relation}).
\item[Table 3] The cutoff $\Lambda$ obtained at the reference point III
for $\tan\beta=0.6,~1.0,~3.0$.
The value of $M_A$ is determined once
we fix the point on $M_{H^{\pm}} - M_h$ plane by
the relation (\ref{eq:relation}).
\item[Table 4] The cutoff $\Lambda$ in the case with the custodial
symmetry in the Higgs potential at the reference point II.
The upper ( lower ) values in each point
correspond to $M_{H^{\pm}}=M_h$ with $\tan\beta=1.0$
( $M_{H^{\pm}}=M_A$ with $\tan\beta$ being free ). GeV unit is used
for the masses of new particles. $M_{i=j}$ means $M_i=M_j$.
\item[Table 5] The cutoff $\Lambda$ in the case with the custodial
symmetry in the Higgs potential at the reference point III.
The upper ( lower ) values in each point
correspond to $M_{H^{\pm}}=M_h$ with $\tan\beta=1.0$
( $M_{H^{\pm}}=M_A$ with $\tan\beta$ being free ).
GeV unit is used for new particles. $M_{i=j}$ means $M_i=M_j$.
\item[Table 6]
The cutoff $\Lambda$ for $M_H=80, 60, 40, 20$ GeV at each value of $M_t=175$
GeV and $125$ GeV. The values in parentheses correspond to the case for
$M_t=125$ GeV. $\lambda_{1=2=3}$ and $\lambda_{4=5}$ mean
$\lambda_1=\lambda_2=\lambda_3$ and $\lambda_4=\lambda_5$,
respectively.
\end{itemize}
\vspace{15pt}
\newpage
\begin{center}
\Large{{\bf Table 1}}
\end{center}
\smallskip
\centering
\begin{tabular}{|c|c|c|l|}\hline
point & $(M_h, M_A)$(GeV) & $\tan\beta$ & $\Lambda$ (TeV) \\ \hline
     &                 & 0.6       & 0.52~;~(0.115)                 \\
A    & (150, 484)          &  1.0       &0.65~;~0.60                  \\
     &               &  3.0       & 0.55~;~(0.057)                 \\ \hline

     &                 & 0.6       & 0.61~;~(0.121)                 \\
B     & (150, 474)          &  1.0       &0.79~;~0.72                  \\
     &               &  3.0       & 0.63~;~(0.058)                 \\ \hline

     &                 & 0.6       & 0.83~;~(0.134)                 \\
C     & (150, 453)          &  1.0       &1.14~;~1.02                  \\
     &               &  3.0       & 0.77~;~(0.061)                 \\ \hline

     &                 & 0.6       & 1.65~;~(0.168)                 \\
D     & (150, 410)          &  1.0       &2.45~;~2.17                  \\
     &               &  3.0       & 0.95~;~(0.067)                \\ \hline

     &                 & 0.6       & 0.61~;~(0.122)                 \\
E     & (200, 471)          &  1.0       &0.83~;~0.76                 \\
     &                  & 3.0       &(0.32)~;~(0.059)               \\\hline
     &               &  0.6       & 0.83~;~(0.136)                 \\
F     & (200, 449)    &  1.0      &1.22~;~1.10                  \\
     &               &  3.0      &(0.33)~;~(0.061)            \\\hline
     &               &  0.6       & 1.57~;~(0.177)                \\
G     & (200, 406)    &  1.0       &2.71~;~2.40                  \\
     &              & 3.0        &(0.33)~;~(0.068)         \\\hline
     &               &  0.6       & 0.97~;~(0.151)                 \\
H     & (240, 431)          &  1.0       &1.78~;~1.57                  \\
     &                &  3.0         & (0.18)~;~(0.064)      \\\hline
\end{tabular}
\newpage
\begin{center}
\Large{{\bf Table 2}}
\end{center}
\smallskip
\centering
\begin{tabular}{|c|c|c|c|}\hline
point & $(M_h, M_A)$(GeV) & $\tan\beta$ & $\Lambda$ (TeV) \\ \hline
     &                 & 0.6       & 0.65                 \\
B     & (150, 472)      &  1.0       &0.81                  \\
     &               &  3.0       & 0.64                 \\ \hline

     &                 & 0.6       & 2.52                 \\
C     & (150, 389)          &  1.0       &3.95                  \\
     &               &  3.0       & 1.06                 \\ \hline

     &                 & 0.6       & 0.78                 \\
E     & (200, 458)    &  1.0       &1.09                  \\
    &                & 3.0        & (0.328)             \\ \hline

     &                 & 0.6       & 1.12                 \\
F     & (200, 435)   &  1.0       &1.75                  \\
     &              & 3.0         & (0.334)             \\\hline

     &                 & 0.6       & 2.14                 \\
G     &(200, 385)    &  1.0       &4.4                  \\
     &              &   3.0       & (0.338)             \\ \hline

     &                 & 0.6       & 0.99                 \\
I     &(250, 427)    &  1.0       &2.08                  \\
     &               &  3.0       & (0.162)             \\\hline

     &                 & 0.6       & 1.22                 \\
J     & (250, 374)   &  1.0       &5.25                  \\
     &               &  3.0          & (0.162)             \\\hline

     &                 & 0.6       & 0.55                 \\
L     &(300, 352)    &  1.0       &6.41                  \\
     &               &  3.0       & (0.106)             \\\hline

     &              & 0.6        & (0.29)              \\
M     &(352, 300)   &  1.0       &5.93                  \\
     &              & 3.0         & (0.080)            \\\hline

    &               & 0.6         & (0.16)            \\
O   & (427, 250)   & 1.0        &1.75                \\
   &                 & 3.0       & (0.064)            \\ \hline

    &               & 0.6         & (0.13)            \\
P   & (458, 200)   & 1.0        &0.97                \\
   &                 & 3.0       & (0.061)            \\ \hline

    &               & 0.6         & (0.12)            \\
Q   & (472, 150)   & 1.0        &0.74                \\
   &                 & 3.0       & (0.058)           \\ \hline
\end{tabular}
\newpage
\begin{center}
\Large{{\bf Table 3}}
\end{center}
\smallskip
\centering
\begin{tabular}{|c|c|c|c|}\hline
point & $(M_h,M_A)$(GeV) & $\tan\beta$ & $\Lambda$ (TeV) \\ \hline
     &                 & 0.6       & 3.55                 \\
B     & (150, 357)     &  1.0       &6.02                  \\
     &               &  3.0       & 1.14                 \\ \hline

     &                 & 0.6       & 0.73                 \\
C     & (200, 466)    &  1.0       &0.95                  \\
     &               &  3.0       & (0.328)                 \\ \hline

     &                 & 0.6       & 2.56                 \\
E     & (200, 351)    &  1.0       &6.48                  \\
    &                & 3.0       & (0.343)                      \\\hline

     &                 & 0.6       & 0.77                 \\
F     & (250, 456)   &  1.0       &1.19                \\
    &                & 3.0       & (0.162)                      \\\hline

     &                 & 0.6       & 1.29                 \\
H     & (250, 337)    &  1.0       &7.12                  \\
    &                & 3.0       & (0.155)                       \\\hline

     &                 & 0.6       & 0.55                \\
I     & (300, 431)   &  1.0       &2.09                 \\
    &                & 3.0       & (0.105)                       \\\hline

     &                 & 0.6       & 0.59                \\
K     & (300, 333)   &  1.0       &8.40                 \\
    &                & 3.0       &(0.106)                       \\\hline

     &                 & 0.6       &(0.310)                \\
L     & (350, 407)   &  1.0       &3.44                 \\
    &                & 3.0       &(0.081)                       \\\hline

     &                 & 0.6       & (0.313)                \\
M     & (350, 372)   &  1.0       &6.61                 \\
    &                & 3.0       &(0.081)                       \\\hline

     &                 & 0.6       & (0.199)                \\
O     & (400, 361)   &  1.0       &3.20                 \\
    &                & 3.0       &(0.068)                       \\\hline

     &                 & 0.6       & (0.199)                \\
P     & (400, 304)   &  1.0       &3.46                \\
    &                & 3.0       &(0.068)                       \\\hline

     &                 & 0.6       & (0.152)               \\
Q     & (440, 297)   &  1.0       &1.45                 \\
    &                & 3.0       &(0.062)                 \\\hline
\end{tabular}
\newpage
\begin{center}
\Large{{\bf Table 4}}
\end{center}
\smallskip
\centering
\begin{tabular}{|c|c|c|c|}\hline
point & $(M_{H^{\pm}=h}, M_A)$           & $\tan\beta$ & $\Lambda$ (TeV) \\
      & $(M_{H^{\pm}=A}, M_h)$           &            &             \\ \hline

A     & (150, 477)          &  1.0       &0.73             \\ \cline{2-4}
     &                     & 0.6        & (0.121)        \\
    &  (150, 477)         & 1.0        & 0.67             \\
    &                     & 3.0         & (0.058)        \\\hline

D     & (200, 469)          &  1.0       &0.86             \\\cline{2-4}
     &                     & 0.6        & (0.126)        \\
    &  (200, 469)         & 1.0        & 2.44             \\
    &                     & 3.0         & (0.059)        \\\hline

H    & (250, 452)          &  1.0       &1.24             \\ \cline{2-4}
     &                     & 0.6        & (0.138)        \\
    &  (250, 452)         & 1.0        & 1.09             \\
    &                     & 3.0         & (0.061)        \\\hline

K     & (300, 413)          &  1.0       &2.86             \\\cline{2-4}
     &                     & 0.6        & (0.138)        \\
    &  (300, 413)         & 1.0        & 2.35             \\
    &                     & 3.0         & (0.061)       \\\hline

N     & (350, 303)          &  1.0       &6.03             \\\cline{2-4}
     &                     & 0.6        & 0.53        \\
    &  (350, 303)         & 1.0        & 6.47             \\
    &                     & 3.0         & (0.103)        \\\hline
\end{tabular}
\newpage
\begin{center}
\Large{{\bf Table 5}}
\end{center}
\smallskip
\centering
\begin{tabular}{|c|c|c|c|}\hline
point & $(M_{H^{\pm}=h}, M_A)$           & $\tan\beta$ & $\Lambda$ (TeV) \\
      & $(M_{H^{\pm}=A}, M_h)$           &            &             \\ \hline
A     & (150, 470)          &  1.0       &0.86             \\\cline{2-4}
     &                     & 0.6        & (0.113)        \\
    &  (150, 470)         & 1.0        & 0.78             \\
    &                     & 3.0         & (0.058)        \\\hline

D     & (200, 462)          &  1.0       &1.04             \\\cline{2-4}
     &                     & 0.6        & (0.135)        \\
     &  (200, 462)         & 1.0        & 0.94             \\
    &                     & 3.0         & (0.059)        \\\hline

G    & (250, 443)          &  1.0       &1.58             \\ \cline{2-4}
     &                     & 0.6        & (0.149)        \\
    &  (250, 443)         & 1.0        & 1.38             \\
    &                     & 3.0         & (0.062)        \\\hline

J     & (300, 402)          &  1.0       &4.19             \\ \cline{2-4}
     &                     & 0.6        & (0.197)        \\
    &  (300, 402)         & 1.0        & 3.34             \\
    &                     & 3.0         & (0.068)        \\\hline

N     & (350, 270)          &  1.0       &7.07             \\ \cline{2-4}
     &                     & 0.6        & 0.93        \\
    &  (350, 270)         & 1.0        & 7.83             \\
    &                     & 3.0         & (0.133)        \\\hline
\end{tabular}
\newpage
\begin{center}
\Large{{\bf Table 6}}
\end{center}
\smallskip
\centering
\begin{tabular}{|c|c|c|c|}\hline
$M_H$ (GeV)& $M_{\rm new}$ (GeV)& $\lambda_{1=2=3}=-\lambda_{4=5}$ &
$\Lambda$ (GeV)   \\ \hline
80 &319~(301)& 1.68~(1.50) & 1.45$\times 10^4$~(3.17$\times 10^4$) \\ \hline
60 &289~(264)& 1.38~(1.15) & 6.45$\times 10^4$~(2.74$\times 10^5$) \\ \hline
40 &260~(224)& 1.12~(0.83) & 5.19$\times 10^5$~(1.80$\times 10^7$) \\ \hline
20 &237~(182)& 0.93~(0.54) & 5.02$\times 10^6$~(1.50$\times 10^{11}$)\\ \hline
\end{tabular}

\begin{thebibliography}{99}
\bibitem{weinberg} S.~Weinberg, \PLB{82}{79}{387}.
\bibitem{coleman} S.~Coleman and E.~Weinberg, \PRD{7}{73}{1888}.
\bibitem{takeuchi} M.~Peskin and T.~Takeuchi, \PRL{65}{90}{964};
\PRD{46}{92}{381}.
\bibitem{takenaga} K.~Takenaga, \PTP{92}{94}{987}.
\bibitem{IKN} K.~Inoue, A.~Kakuto, Y.~Nakano, \PTP{63}{80}{234}.
\bibitem{FKM} C.~D.~Froggatt, I.~G.~Knowles and R.~G.~Moorhouse,
\PLB{249}{90}{273}.
\bibitem{FKT} K.~Funakubo, A.~Kakuto and K.~Takenaga, \PTP{91}{94}{341}.
\bibitem{gildner} E.~Gildner and S.~Weinberg, \PRD{15}{76}{3333}.
\bibitem{pomarol} A.~Pomarol and R.~Vega, \NPB{413}{94}{3}.
\bibitem{barbieri} R.~Barbieri, Talk given at the Recontres de la
                   Vallee d'Aoste, La Thuile, Italy, march 1994.
\bibitem{ABC} G.~Altarelli, R.~Barbieri and F.~Caravaglios, \NPB{405}{93}{3}.
\end{thebibliography}
\end{document}